\documentclass[a4paper]{article}
\usepackage{graphicx} 
\usepackage[utf8]{inputenc}
\usepackage[english]{babel}
\usepackage{amsmath}
\usepackage{amsfonts}
\usepackage{amssymb}
\usepackage{geometry}

\usepackage{siunitx}
\DeclareSIUnit\angstrom{\text{\AA}}
\DeclareSIUnit\bar{bar}
\usepackage{cuted}
\usepackage{comment}
\DeclareUnicodeCharacter{0303}{\~{}}
\usepackage[hyphens]{url}
\usepackage[hidelinks]{hyperref}
\usepackage{orcidlink}
\usepackage{float}
\usepackage{bm}
\usepackage{relsize}
\usepackage{multirow}
\usepackage{csquotes}
\usepackage{subfig}
\usepackage{physics}
\usepackage{booktabs}
\usepackage{import}
\usepackage{bbold}
\usepackage[backend=biber, sorting=none, maxbibnames=999, style=numeric-comp]{biblatex}
\usepackage{xcolor}
\usepackage{cleveref}
\usepackage{authblk}
\usepackage{abstract}
\usepackage{scrlayer-scrpage}
\bibliography{bibliography.bib}

\title{Reactor operation induced thermal effects on neutron flux measurements using ${}^3$He neutron detectors at a {TRIGA Mk II} research reactor}
\author[a,$\dagger$]{Sebastian Dorer \orcidlink{0009-0001-1670-5780}}
\author[a, b, $\dagger$]{Clemens Trunner \orcidlink{0009-0007-1800-8667}}
\author[b]{Robert Bergmann}
\author[a, b]{Dieter Hainz}
\author[a]{Erwin Jericha \orcidlink{0000-0002-8663-0526}}
\author[a]{Stephan Sponar \orcidlink{0000-0002-6568-6045}}
\author[b]{Thomas Stummer}
\author[b]{Mario Villa}
\affil[a]{Atominstitut, Technische Universität Wien, 1020 Vienna, Austria}
\affil[b]{TRIGA Center Atominstitut, 1020 Vienna, Austria}
\newpairofpagestyles{firstpage}{
  \clearpairofpagestyles
  \ifoot{\textsuperscript{$\dagger$}\,\textnormal{Corresponding authors:
  sebastian.dorer@tuwien.ac.at,
  clemens.trunner@tuwien.ac.at}}
}
\ohead{}

\date{June 2026}

\begin{document}

\twocolumn[
\maketitle

\begin{onecolabstract}
    A variation of the measured neutron flux using ${}^3$He detectors at a TRIGA Mk II nuclear research reactor during steady-state operation is reported in this article. 
    The observed effect shows a statistically significant anti-correlation ($> 7\,\sigma$) between the temperature in the reactor pool and recorded neutron count rates. 
    Following reactor start-up under nominal operating conditions, the effect occurs for a specific time period until both thermal equilibrium and a constant neutron count rate are reached. 
    Simultaneous neutron and temperature measurements are performed in different configurations around the reactor in order to gain qualitative insights about the temperature dependent count rate behavior. 
\end{onecolabstract}
]

\thispagestyle{firstpage}

\section{Introduction}
\label{sec:introduction}
TRIGA reactors have been widely employed for training and research utilizing both in-core facilities and beamtube experiments \cite{iaea:triga}. Research activities concerning their technical parameters and characteristics are ever ongoing by modeling and experiment. They deal with quantities like temperature and power distribution \cite{Mesquita2023, Birri2024}, in-core neutron flux distribution \cite{RR-Benchmark-TRIGA,Neutron-Flux-Distribution-TRIGA,ZEROVNIK201527,SNOJ2011136,GORICANEC2018407}, neutronics \cite{Melo2024, Prasetya2025,TAN20191715}, reactor critical parameters \cite{Islam2025a} and neutron beam characteristics \cite{Pasman2026}. A variety of different model codes have been developed and used, like TRIPOLI-4 in combination with MCNP \cite{Ghninou2023}, Serpent \cite{Melo2024}, OpenMC \cite{Mira2024, Prasetya2025, Islam2025a}, COOLOD-N2 in combination with MCNP \cite{Trianti2024}, OpenTHY \cite{Ooulad2025}, using steady-state single channel models \cite{Islam2025b} and CFD analysis \cite{Ghazanfari2026, Umar2026,CASTAGNA2022108704}, and advanced reactor dynamics with model order reduction \cite{Introini2024}. These exemplary works clearly show that a series of models and software codes are available to characterize the operational characteristics of the TRIGA reactors. Also several aspects of the TRIGA Mk II nuclear research reactor at the Atominstitut of TU Wien (ATI) and the TRIGA Center Atominstitut have been modeled by MCNP \cite{Khan2011a, Khan2011b, Stummer2013, Cagnazzo2014}, ORIGEN2 \cite{Khan2013} and SuperMC \cite{Khan2019}. Specifically the neutron flux at several positions inside this reactor and its experimental facilities were investigated in \cite{Khan2011a, Khan2011b}.

In this work we want to put emphasis on the measured neutron flux at several experimental stations of the TRIGA Mk II research reactor in Vienna, and specifically describe its temporal behavior. 
A lot of the available literature regarding the modeling of TRIGA Mk II reactors describes the correlation of simulations and measurements during steady-state operation (e.g. \cite{CAMMI2016308,Ooulad2025,Islam2025b, Khan2011b}), without investigating the temporal evolution of the system. 
Therefore, this work offers novel insights into a time-dependent effect occurring after the reactor start-up. 

Thereby, during neutron count rate measurements over the period of a full reactor day, \SI{7}{\hour}, a statistically significant ($> 7\,\sigma$) reduction of the count rate of up to \SI{3.697 \pm 0.006}{\percent} (depending on the measurement location and operating conditions) at nominal \SI{250}{\kilo\watt} operation over a duration of about \SIrange{2.0}{2.5}{\hour} after the reactor start-up was observed. 
Following the first observation, neutron count rate measurements at different beam tubes around the reactor were performed, in order to investigate the reduction effect in more detail. 
Furthermore, after contemplating about plausible explanations for the behavior, temperature data inside the reactor pool was taken simultaneously, leading to the discovery of the apparent anti-correlation between the count rate and temperature, with the count rate decreasing with increasing temperature. 

The aim of the paper is the description of the observed effect, to present already performed measurements and their results, as well as providing suggestions for future experiments in order to further investigate the root cause of the effect.
Our first results indicate the possible requirement of the correction of this effect for high precision experiments performed at TRIGA Mk II reactors (examples are provided in \Cref{sec:results}).

\section{Experimental Setup}
\label{sec:experimental_setup}
\subsection{TRIGA Mk II Reactor}
The TRIGA reactor was invented and designed by General Atomic. It originated with the TRIGA Mk I reactor, following the idea of building a reactor that \enquote{could be given to a bunch of high school children to play with, without any fear that they would get hurt}\cite{physics:dyson}. This objective was achieved through the use of Uranium-zirconium hydride (\emph{U-ZrH}) as nuclear fuel. 
The inherent TRIGA safety derives from a strong negative temperature coefficient. This behavior results of the tight temperature coupling between uranium and hydrogen in the \emph{U-ZrH} fuel, with hydrogen providing a substantial share of neutron moderation. Consequently, thermal power is directly tied to the system’s moderation capacity and self-limits as temperature rises. To improve neutron economy, the core is surrounded by a graphite reflector. Both the core and the reflector are immersed in a water pool, which not only removes heat but also acts as an additional moderator.
The TRIGA Mk II design is based on its predecessor and extends it by incorporating a graphite thermal column and neutron beam ports penetrating the graphite reflector. In the Mk II design, the graphite reflector fulfills several functions already present in the Mk I, primarily the reduction of neutron losses. This is particularly important with respect to the power density resulting from the special fuel type and the associated safety requirements. A further, newly introduced function of the graphite reflector in the TRIGA Mk II design is its use as an additional moderator for the beam tubes.

\subsection{TRIGA Mark II Vienna}

The TRIGA reactor located in Vienna and operated by the TRIGA Center
Atominstitut of TU Wien is a TRIGA Mark II design. The reactor pool has a diameter of \SI{1.98}{\meter} and a depth of
\SI{6.2}{\meter}. The graphite reflector is located \SI{0.61}{\meter}
above the bottom of the reactor tank.

The reflector has a cylindrical shape with an outer diameter of
\SI{1.090}{\meter}, an inner diameter of \SI{0.457}{\meter}, and a height
of \SI{0.560}{\meter}.

\begin{figure*} 
    \centering
    \includegraphics[width=0.7\textwidth]{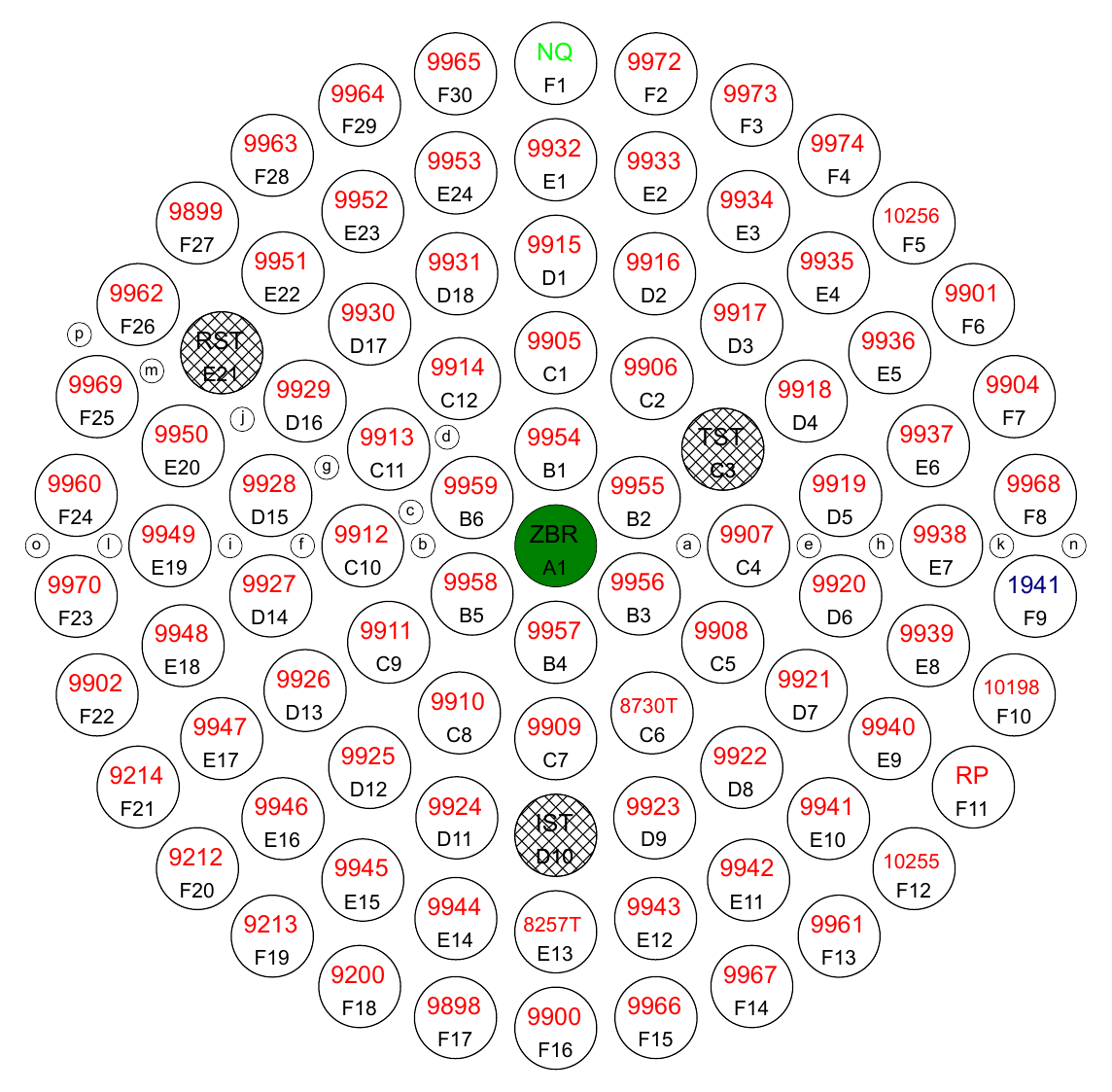}
    \caption{Reactor core configuration of the TRIGA Mark II reactor at TU Wien at the time of the measurements. Legend: red numbers indicate fuel elements, the blue number indicates a graphite dummy element, NQ the neutron source, ZBR the central irradiation position, RP the pneumatic rabbit system, RST the regulating rod, IST the transient rod, and TST the safety rod. B1--F30 are position markers, and a--p dedicated in-core neutron flux measurement positions.}
    \label{fig:Reactor_Core}
\end{figure*}

The reactor core is located inside the reflector and consists of
84 fuel elements, as shown in \Cref{fig:Reactor_Core}. Of these,
82 fuel elements are Type~104 fuel elements containing
\SI{8.5}{\percent} uranium by weight with an enrichment of
\SI{20}{\percent} in \textsuperscript{235}U. This corresponds to
\SI{38.19}{\gram} of \textsuperscript{235}U per fuel element.
Two fuel elements (8257 and 8730) are instrumented fuel elements of
Type~204 with the same uranium content.

Since the new reactor core was implemented in 2012, a total energy of \SI{2632}{\mega\watt\hour} has been released until 27.03.2026. 
This corresponds to an estimated burnup of approximately 6.84 MW$\,$d$\,$kg$_{\text{HM}}^{-1}$.

\begin{figure*}
    \centering
    \includegraphics[width=0.7\textwidth]{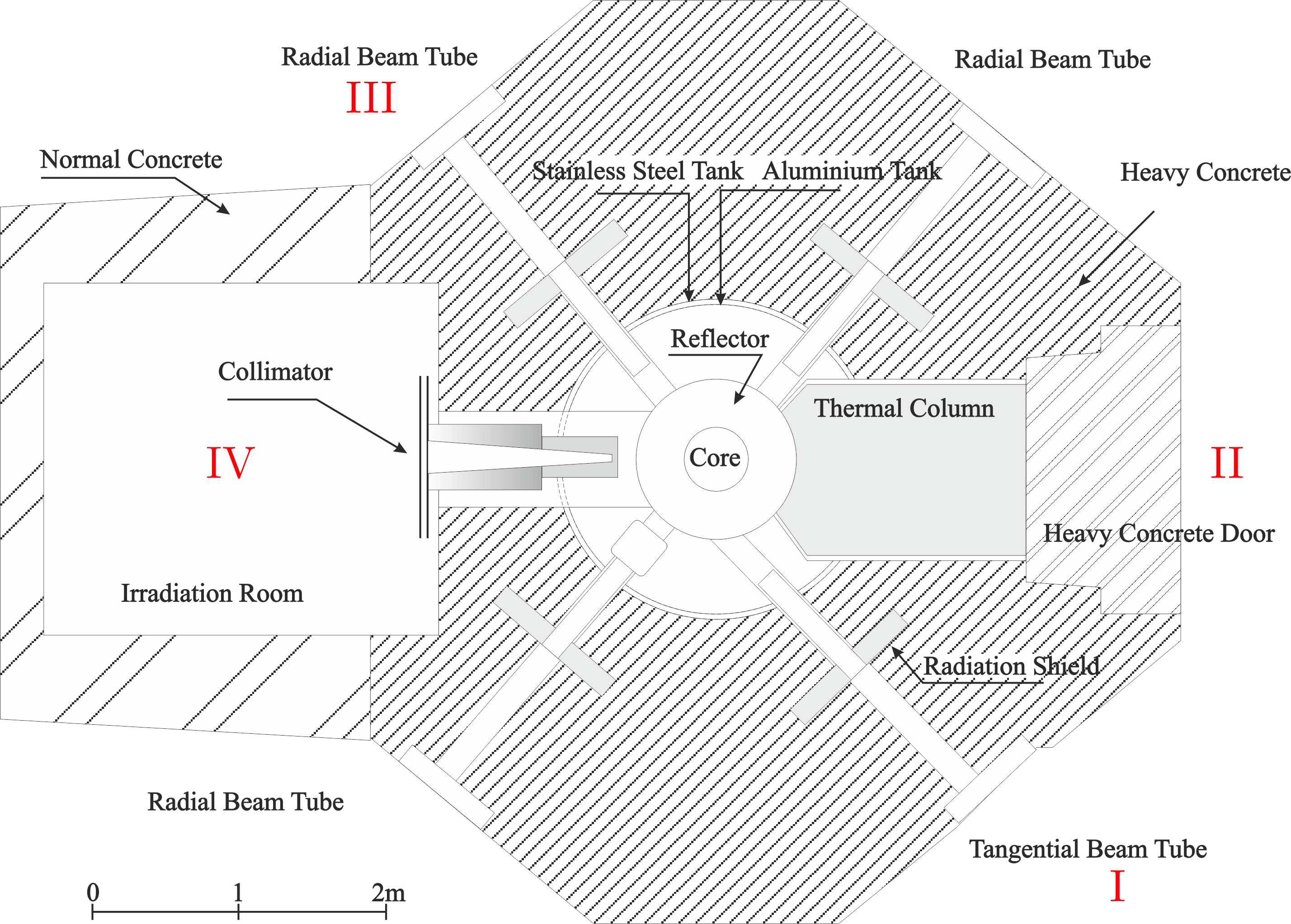}
    \caption{Schematic top-down drawing of the TRIGA Mk II nuclear research reactor layout. The measurements described in this paper are performed at the tangential beam tube (\color{red}I\color{black}), thermal column (\color{red}II\color{black}), radial beam tube (\color{red}III\color{black}) and dry irradiation room (\color{red}IV\color{black}).}
    \label{fig:Reactor_Horizontal}
\end{figure*}

As shown in  \Cref{fig:Reactor_Horizontal}, the four beam tubes, as well as the beam port for the dry irradiation room and the thermal column, terminate in or at the graphite reflector depending on the desired neutron spectrum. In particular, the tangential beam tube passing the core receives its neutrons via scattering processes in the graphite reflector.
To realize the concept of a reactor suitable for training, research, and isotope production, both the TRIGA Mk I and Mk II employ an open reactor pool with natural convection cooling through the core. The deposited thermal energy is removed by extracting pool water from the bottom of the tank, cooling the water, and reinserting it at approximately half the height of the pool \cite{iaea:triga}.
This primary cooling circuit provides a flow rate of \SI{27}{\cubic\meter\per\hour}. Heat is transferred to the environment through a secondary cooling circuit via a plate heat exchanger with a rated thermal power of \SI{1}{\mega\watt}. The secondary circuit has a flow rate of \SI{18.5}{\cubic\meter\per\hour}. Both cooling circuits are in use during reactor operation. In addition, a dedicated purification circuit is operated continuously, with a flow rate of \SI{2.8}{\cubic\meter\per\hour}, to maintain the required water quality.

\subsection{Neutron Detectors}
Almost all neutron measurements presented in this paper were performed using a \emph{VacuTec type 70 063} ${}^3$He-filled gas proportional neutron detector with a diameter of \SI{1}{inch} and an effective length of \SI{10}{\centi\meter} \cite{vacutec:he3_flyer, vacutec:he3_application}. 
It has a nominal ${}^3$He gas pressure of \SI{4}{\bar} and is operated at \SI{760}{\volt}. 
The data at the dry irradiation room was taken with a \emph{Canberra 24NH15} ${}^3$He-filled (\SI{4}{\bar}) \autocite{canberra:he3_catalog} neutron counter operated at \SI{900}{\volt}. 
The detectors' output was connected to a \emph{mesytec MRS-2000} preamplifier and pulse shaper \cite{mesytec:mrs2000}. 
This voltage signal was routed into an \emph{Amptek MCA-8000D} multichannel analyzer \cite{amptek:mca} controlled via the \emph{Amptek DPPMCA} data acquisition software \cite{amptek:dppmca}. 

\subsection{Temperature Probes}
Temperature measurements were carried out at different measurement points using different systems. The main temperature measurement was performed using a \emph{Calex Excelog 6} temperature data logger \cite{calex:excelog6} in combination with \emph{RS PRO} Type K thermocouples  \cite{rs:thermocouple}. The thermocouples have a cable length of \SI{10}{\meter}, an insulated probe diameter of \SI{1}{\milli\meter} and  \SI{0.3}{\milli\meter} wire gauge.
\emph{Calex} specifies the measurement accuracy for the \emph{Excelog 6} with $\pm$ \SI{0.8}{\celsius} below \SI{800}{\celsius} for thermocouples. The thermocouples themselves have a typical systematic deviation from \SIrange{0.5}{5}{\celsius}. 
To obtain the graphite temperature as accurately as possible, two thermocouples were positioned on the reflector cask using weights made of aluminum. Sensor one was placed in the reflector groove originally intended for the rotary specimen rack, while sensor two was positioned on top of the reflector.

As an additional source of temperature data, the reactor’s I\&C system was used. The reactor pool is equipped with a PT100 temperature sensor to measure the water temperature. This sensor is located \SI{50}{\centi\meter} below the water surface, several meters away from the reactor core.

\section{Measurements}
\label{sec:measurements}
To investigate the cause of the decreasing neutron count rate, measurements were performed with different experimental configurations. 
The reactor was operated at power levels of \SI{5}{\kilo\watt}, \SI{100}{\kilo\watt} and \SI{250}{\kilo\watt}. 
As soon as the requested power level was reached, the reactor instrumentation was switched to its automatic control mode, regulating the control rod position to keep the power constant. 
During reactor operation, the primary water cooling circle was active for all measurements except for two specific runs at \SI{5}{\kilo\watt} and \SI{100}{\kilo\watt} without cooling. 
Measurements were performed at the tangential beam tube (\Cref{fig:CRAB-beam}), a radial beam tube (\Cref{fig:White_beam_250kW_with_cooling}), the irradiation room (\Cref{fig:TBS}) and thermal column (\Cref{fig:Radiography_250kW_with_cooling}), therefore allowing to investigate if the observed reduction effect is specific to one beam tube or a global effect observable at all measurement positions. 
\Cref{fig:Reactor_Horizontal} provides a reference for the different measurement locations around the reactor. 
For the runs at the tangential beam tube and irradiation room (except at \SI{5}{\kilo\watt}), the initial neutron beam is reflected once on a highly oriented pyrolytic graphite (HOPG) crystal acting as a monochromator. 
At the irradiation room (for measurements at \SI{250}{\kilo\watt}), a two-layer array of 2 HOPG crystals with dimensions of $\SI{50}{\milli\metre} \times \SI{50}{\milli\metre} \times \SI{5}{\milli\metre}$ each in the first layer, and eight $\SI{20}{\milli\metre} \times \SI{20}{\milli\metre} \times \SI{2}{\milli\metre}$ crystals in the second layer on-top, was positioned at a Bragg angle of $\theta_\text{B} \approx \SI{45}{\degree}$ to reflect the initial beam to the detector. 
This intermediate step is necessary at the irradiation room for measurements at \SI{250}{\kilo\watt}, as the initial neutron flux would saturate the available neutron detectors. 
A grade ZYB HOPG crystal with dimensions of $\SI{50}{\milli\metre} \times \SI{50}{\milli\metre} \times \SI{4}{\milli\metre}$ is permanently positioned at approximately \SI{30}{\centi\metre} from the tangential beam tube (location I in \Cref{fig:Reactor_Horizontal}) with $\theta_\text{B} = \SI{23.17}{\degree}$. 
In the case of the tangential beam tube, there is a HOPG monochromator setup consisting of three HOPG crystals, rotated at different Bragg angles, installed at the beam tube exit, in order to supply three experiments from one initial beam tube. 
For all measurements at the tangential beam tube described in this paper, the count rate in the reflected neutron beam was measured. 
Therefore, only thermal neutrons of certain wavelengths fulfilling Bragg's reflection condition \autocite{bragg:reflection-by-crystals} are reflected towards the neutron detector, e.g. \qtylist{0.88; 1.32; 2.64}{\angstrom} at the tangential beam \autocite{Abele2025}. 

During each run, multiple consecutive neutron count rate and temperature measurements were taken. 
Neutron counts were acquired for 60 seconds per measurement and the temperature sensors were read out every 60 seconds. 

The fuel temperature was also measured at six different positions. These temperatures are practically constant at a given reactor power from the reactor start-up until the shut-down. There is no obvious correlation between the fuel temperature and the neutron count rate.

\section{Results}
\label{sec:results}
The measurements under nominal reactor operation conditions (\SI{250}{\kilo\watt} and active cooling) show that the reduction of the count rate occurs on a time scale of about \SIrange{2}{2.5}{\hour} after the reactor start-up. 
This time scale introduces an upper limit to the duration of individual measurements during one run of about \SI{10}{\minute}, to still provide sufficient temporal resolution to observe the reduction effect. 
However, in order to achieve the desired statistical uncertainty of less than 1\%, the count rate per time interval must still be high enough. 
Therefore, a tradeoff exists between the temporal resolution of the reduction effect and statistical uncertainty per time bin. 
Ideally, for more precise investigations of the effect, these measurements should be performed with detectors with large active detection volumes, high detection efficiency and low dead-time.
The influence of the reduction effect on the integral of the counts over a whole reactor day is less than 1\%, however, due to the percent-level change of the count rate from the reactor start-up to the plateau region over a span of less than \SI{2.5}{\hour}, it may have consequences for high precision experiments relying on stable conditions, e.g. scattering experiments using a scanning technique \autocite{Jericha2003} or neutron interferometry experiments which require stability during the acquisition of the complete interference pattern \autocite{Terburg1993}.

\begin{figure*}
    \centering
    \begin{minipage}[c]{0.49\textwidth}
        \includegraphics[width=1.0\textwidth]{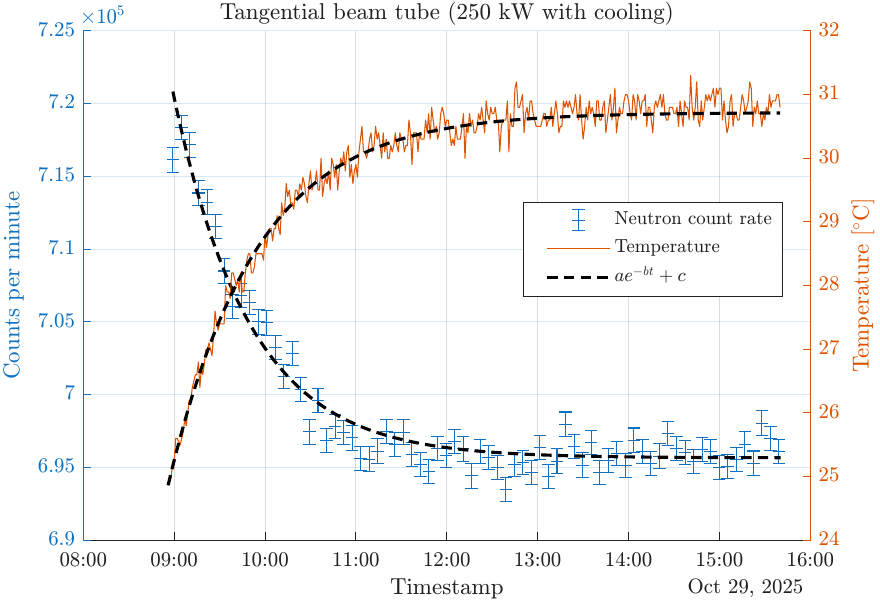}
        \caption*{a)}
    \end{minipage}
    \begin{minipage}[c]{0.49\textwidth}
        \includegraphics[width=1.0\textwidth]{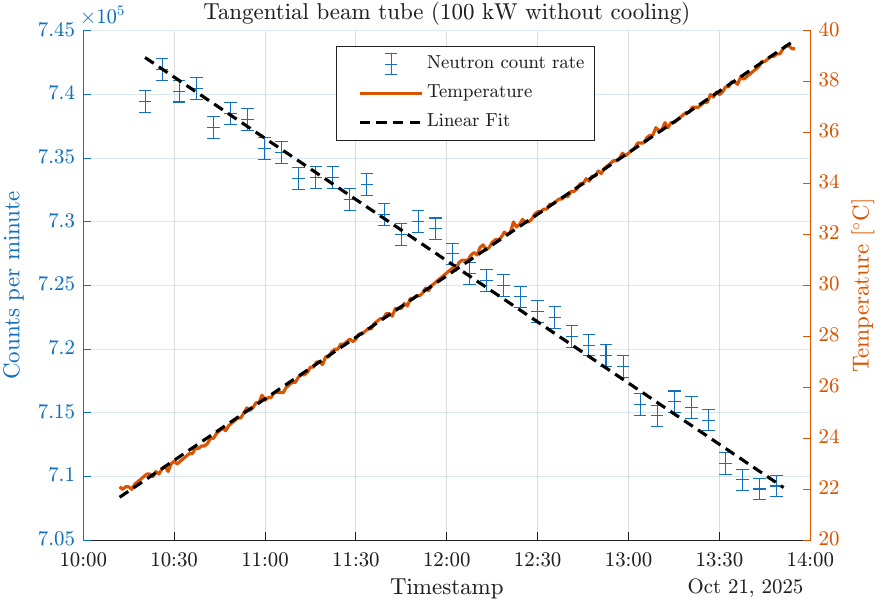}
        \caption*{b)}
    \end{minipage}
    \caption{Measurement position I: Neutron count rate and temperature measurement results at the tangential beam tube. a) During nominal reactor operation at \SI{250}{\kilo\watt} with cooling. Both curves are fitted with $f(t)=ae^{-bt}+c$ and the fit parameters are listed in \Cref{tab:fit-parameter-exp}. b) At \SI{100}{\kilo\watt} with the cooling circuit deactivated. Both curves are fitted with $f(t)=kt+d$. For visual clarity only one count rate data point per five minutes is shown in the plots.}
    \label{fig:CRAB-beam}
\end{figure*}

By comparing the temporal behavior of the neutron count rate and temperature measured by the sensor in the reflector groove for runs with and without cooling (\Cref{fig:CRAB-beam}), the data does not only show a statistically significant decrease of the measured neutron count rate, but also the clear anti-correlation between the two measured quantities. 
For the case without cooling, the calculated Pearson correlation coefficient \autocite{Pearson1896} is $r = -0.9914$, with a p-value of $p = \SI{6.312E-21}{}$, which corresponds to a significance of more than $9 \,\sigma$. 
This hints to a temperature dependency of the count rate reduction. 
Due to the high neutron count rates (at all measurement positions other than the radial beam tube) in the order of $\mathcal{O}(10^4)$ per second and a measurement duration of 60 seconds, the statistical uncertainty from the underlying Poisson process is $\epsilon_\text{stat.} \approx 0.12 \%$, and therefore one order of magnitude smaller than the observed count rate reduction of around $1 \% - 5 \%$ presented in this work. 

The neutron count rate and temperature data has been fitted with $f(t) = ae^{-bt}+c$ for all runs, except the ones at the thermal column, the tangential beam tube with \SI{100}{\kilo\watt}, and the irradiation room with \SI{5}{\kilo\watt}, for which $f(t) = kt + d$ was used. 
The results listed in \Cref{tab:fit-parameter-exp} show that the values of $b$ for different measurement positions are remarkably similar. This further strengthens the assumption that the observed effect is indeed a global reactor effect and not specific to the different beam tubes. 

For the linear fits of the measurements at the tangential beam tube at \SI{100}{\kilo\watt} without cooling, the slopes are $k_n = -1.60(1) \times 10^2$ and $k_T = 8.05(1)\times 10^{-2}$, and the offsets $d_n = 7.429(2) \times 10^5$ and $d_T = 21.69(2)$, for the count rate and temperature, respectively.
The obtained $R^2$-values are $R^2_n = 0.9880$ and $R^2_T = 0.9995$. 

\begin{table*}[htbp!]
    \centering
    \begin{tabular}{cccccc}
        \toprule
        Position & Qty. & $a$ & $b$ $\left[\unit{\per\minute}\right]$ & $c$ $\left[\unit{\per\minute}\right]$,$\left[\unit{\celsius}\right]$ & $R^2$\\
        \midrule
         I & $n$ & $3.68 \cdot 10^4 \pm 7.66 \cdot 10^2$ & $1.98 \cdot 10^{-2} \pm 4.88 \cdot 10^{-4}$ & $6.96 \cdot 10^5 \pm 90.6$ & 0.9545 \\
         III & $n$ & $5.19 \cdot 10^2 \pm 38.8$ & $1.97 \cdot 10^{-2} \pm 2.73 \cdot 10^{-3}$ & $1.79 \cdot 10^{4} \pm 11.9$ & 0.4237 \\
         IV & $n$ & $5.81 \cdot 10^{3} \pm 1.93 \cdot 10^{2}$ & $1.54  \cdot 10^{-2} \pm 9.88  \cdot 10^{-4}$ & $4.07  \cdot 10^{5} \pm 65.6$ & 0.7752 \\
         \midrule
         I${}^\dagger$ & $T$ & $-5.85 \pm 6.16 \cdot 10^{-2}$ & $1.72 \cdot 10^{-2} \pm 3.17 \cdot 10^{-4}$ & $30.7 \pm 1.62 \cdot 10^{-2}$ & 0.9682 \\
         II & $T$ & $-6.86 \cdot 10^{-1} \pm 2.13 \cdot 10^{-2}$ & $2.69 \cdot 10^{-2} \pm 1.35 \cdot 10^{-3}$ & $28.8 \pm 4.03 \cdot 10^{-3}$ & 0.7998 \\
         III & $T$ & $-6.55 \pm 1.85 \cdot 10^{-2}$ & $1.55 \cdot 10^{-2} \pm 8.53 \cdot 10^{-5}$ & $29.5 \pm 6.35 \cdot 10^{-3}$ & 0.9977 \\
         IV & $T$ & $-1.35 \pm 1.50 \cdot 10^{-2}$ & $1.17 \cdot 10^{-2} \pm 2.89 \cdot 10^{-4}$ & $28.9 \pm 6.94 \cdot 10^{-3}$ & 0.9592 \\
         \bottomrule
    \end{tabular}
    \caption{Fit parameters for the exponential fits ($f(t) = ae^{-bt} + c$) of the neutron count rate $n$ and temperature $T$ at the different measurement positions. All measurements were performed at \SI{250}{\kilo\watt}. The temperature of ${}^\dagger$ is recorded with the temperature probes described previously, whereas the other temperature values correspond to the reactor pool temperature measured by the reactor instrumentation. The table includes $\pm\sigma$ confidence intervals of the fit parameters.}
    \label{tab:fit-parameter-exp}
\end{table*}

\begin{figure*}
    \centering
    \begin{minipage}[c]{0.49\textwidth}
        \includegraphics[width=1.0\textwidth]{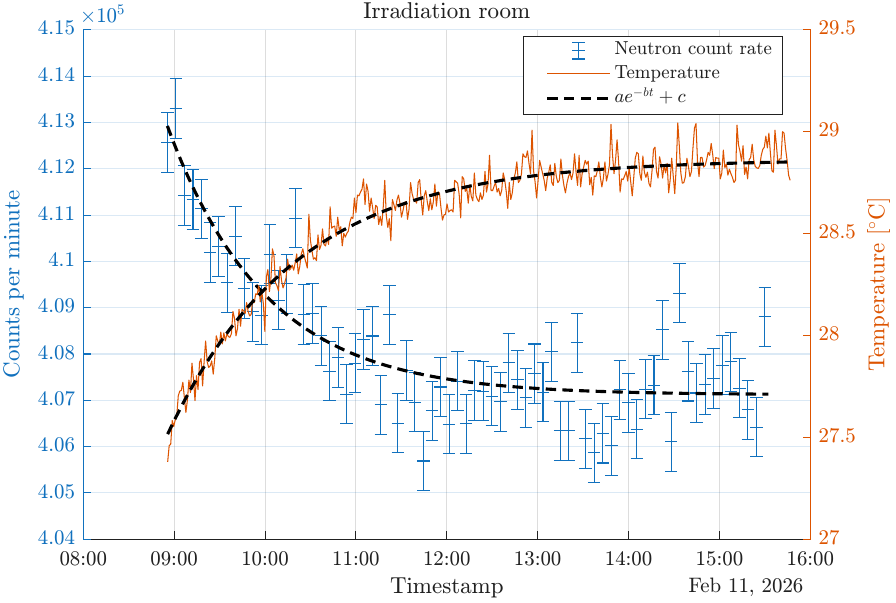}
        \caption*{a)}
    \end{minipage}
    \begin{minipage}[c]{0.49\textwidth}
        \includegraphics[width=1.0\textwidth]{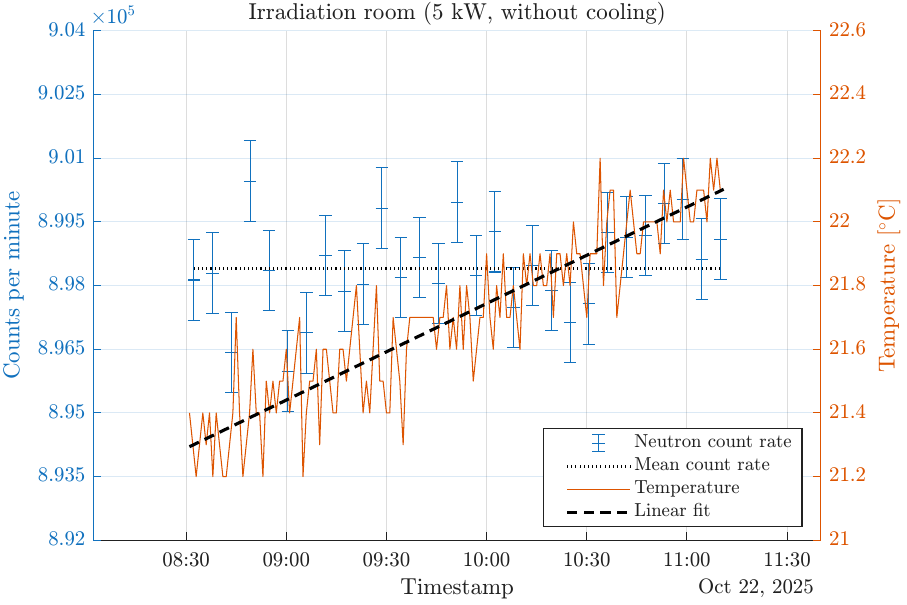}
        \caption*{b)}
    \end{minipage}
    \caption{Measurement position IV: Time dependent neutron count rate measured at the neutron beam at the dry irradiation room. a) Reflected beam during nominal reactor operation at \SI{250}{\kilo\watt} with cooling. The data is fitted with $f(t)=ae^{-bt}+c$ and the fit parameters are listed in \Cref{tab:fit-parameter-exp}. b) Measurement run in the direct neutron beam (not reflected) at \SI{5}{\kilo\watt} without cooling. The dotted black line corresponds to the mean value of the neutron count rate within the recorded time window and the dashed black line to a linear fit of the temperature data. For visual clarity only one count rate data point per five minutes is shown.}
    \label{fig:TBS}
\end{figure*}

\begin{figure}
    \centering
    \includegraphics[width=1\linewidth]{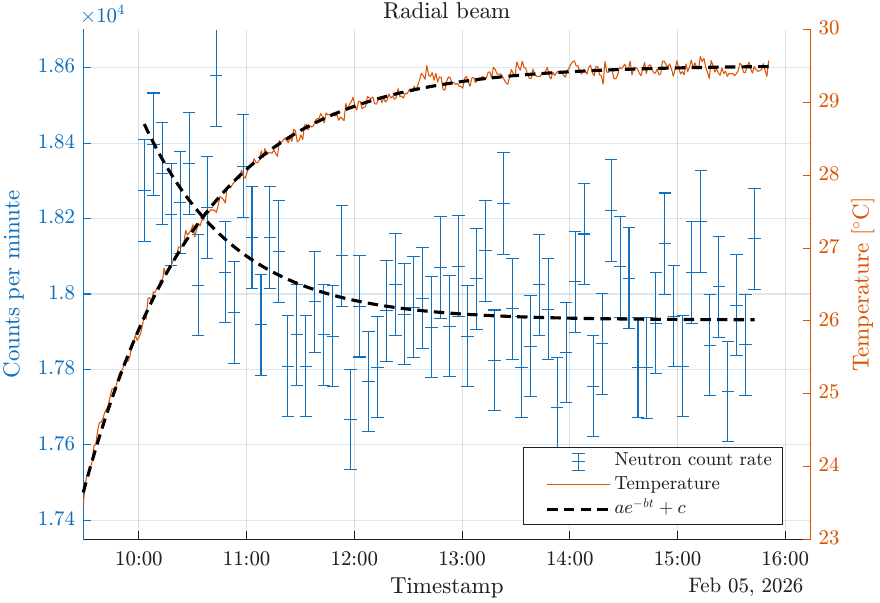}
    \caption{Measurement position III: Time dependent neutron count rate measured at a radial beam tube during nominal reactor operation (\SI{250}{\kilo\watt} with cooling). For visual clarity only one count rate data point every five minutes is shown in the figure.}
    \label{fig:White_beam_250kW_with_cooling}
\end{figure}

\begin{figure}
    \centering
    \includegraphics[width=1\linewidth]{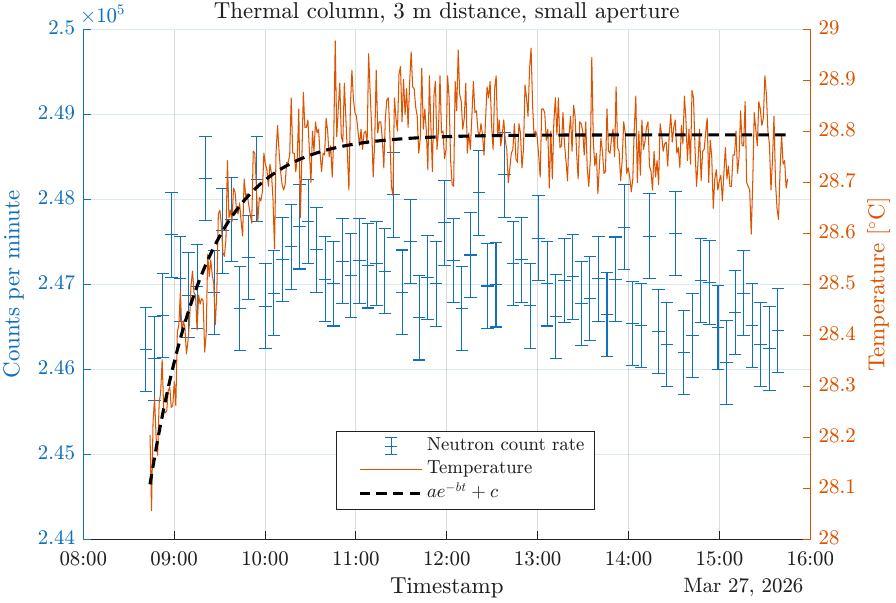}
    \caption{Measurement position II: Time dependent neutron count rate measured at the thermal column at a distance of \SI{3}{\meter} from the reactor wall and with an aperture in front of the detector during nominal reactor operation (\SI{250}{\kilo\watt} with cooling). There is no observable correlation between the water temperature and the neutron count rate ($r = 0.0859$). For visual clarity only one count rate data point every five minutes is shown in the figure.}
    \label{fig:Radiography_250kW_with_cooling}
\end{figure}

Since the count rate reduction described above is observed for measurements at different locations at the reactor (see \Cref{fig:CRAB-beam,fig:White_beam_250kW_with_cooling,fig:TBS}), this strongly hints at an intrinsic reactor-wide effect and not an effect specific to only one beam tube.
For the effect that is observable during nominal operation at the tangential beam tube (\SI{250}{\kilo\watt} with cooling, see \Cref{fig:CRAB-beam}a), the calculated Pearson correlation coefficient is $r = -0.9526$, with a p-value of $p = \SI{7.425E-24}{}$, which corresponds to a significance of $>10 \,\sigma$. 
Performing the same analysis for the radial beam tube (see \Cref{fig:White_beam_250kW_with_cooling}) yields $r = -0.6467$, $p = \SI{8.967E-42}{}$, and a significance $> 13 \, \sigma$.
The thermal column is the only position at which the effect was not observable at \SI{250}{\kilo\watt} (see \Cref{fig:Radiography_250kW_with_cooling}). For this case, the Pearson correlation coefficient is $r = 0.0859$, with a p-value of $p = 0.5657$, corresponding to a significance of $\approx 0.57 \, \sigma$. The absence of the correlation could be explained by the significantly different construction of the thermal column compared with the other beam tubes. The difference lies in the fact that the thermal column is not a beam tube in the conventional sense. Instead, it consists of a graphite-brick assembly directly located at the reflector and has dimensions of approximately $\SI{1}{\meter} \times \SI{1}{\meter} \times \SI{1.7}{\meter}$ \autocite{iaea:triga}. For the experiment, a collimator configuration was used, where the central graphite brick was replaced by a conical collimator \autocite{BuchbergerThomas1986NUaT}. Consequently, only about 35 cm of graphite remain along the beam direction. The combination of this graphite path length and the substantial dimensions of the thermal column enables the neutrons to reach near-thermal equilibrium with the ambient temperature before exiting the collimator, thereby considerably diminishing the effect.
In addition, the observation that the effect does not occur at the thermal column excludes the reactor instrumentation as an origin. 

The measurements at the irradiation room with a reactor power of \SI{250}{\kilo\watt} (see \Cref{fig:TBS}a) yield $r = -0.8718$, $p = \SI{6.490E-15}{}$, and a corresponding significance of $\approx 7.8 \, \sigma$.
At a reactor power of \SI{5}{\kilo\watt} without cooling (see \Cref{fig:TBS}b), the data shows a constant (within the margins of error) neutron count rate and a slight temperature increase of $\Delta T < \SI{1}{\kelvin}$ $\left(\approx \SI{0.0051 \pm 0.0002}{\kelvin\per\minute}\right)$, with $R^2 = 0.8258$. 
Here, the Pearson correlation coefficient is $r = -0.2643$, with a p-value of $p = 0.289$ corresponding to a significance of only $\approx 1 \, \sigma$. 
Combining these observations additionally supports the theory of a temperature dependency of the measured neutron count rate, since the reduction effect occurred at the same beam tube for a larger temperature increase (see \Cref{fig:TBS}a).

\section{Summary}
\label{sec:summary}
Utilizing ${}^3$He neutron detectors, a statistically significant ($> 9\,\sigma$) reduction of the neutron count rate of up to \SI{3.697 \pm 0.006}{\percent} (depending on the measurement location and operating conditions) over a duration of about \SIrange{2.0}{2.5}{\hour} following the reactor start-up was observed during nominal \SI{250}{\kilo\watt} operation at the TRIGA Mk II research reactor at TRIGA Center Atominstitut, TU Wien, Austria. 
Simultaneous temperature measurements inside the reactor pool provide a clear temperature dependence of the rate reduction, with the rate decreasing with increasing temperature. 
Calculating the Pearson correlation coefficient for the relation of the neutron count rate and reactor pool temperature yields a statistically significant ($> 7 \, \sigma$) anti-correlation. 
Furthermore, by performing the measurements at different beam tubes around the reactor, the effect is shown to occur at all beam tubes reactor-wide, with the exception of the thermal column. 

The investigations performed for this paper lead to the conclusion that the described behavior depends on the specific reactor design, characteristics of the installation, power level, operational mode and schedule. 
If the observed behavior is present at all TRIGA Mk II reactors, it could have consequences for high-precision experiments relying on the measurement of relative count rates. 

Since the reactor is a complex coupled thermal system, our current results and experimental limitations do not allow the concrete determination of the effect's origin. 
Searching for options to experimentally decouple contributions of individual parts (e.g. cooling water and graphite reflector) to the temperature effect is critical to further progress. 

\section{Outlook}
\label{sec:outlook}
The authors suggest to reproduce the current results presented in this paper at other TRIGA Mk II reactors. 
This would also provide insights into possible contributions of the reactor instrumentation. 
Additionally, the implementation of neutron monitors for high-precision experiments is recommended, in order to allow for a correction of the effect during data analysis. 

Furthermore, the complex interactions between thermo-fluid-dynamics and neutron physics within the realm of nuclear reactors is probably not limited to small research reactors, but could extend to other reactor types. 
For example, a temperature-dependent, percent-level fluctuation of the thermal neutron flux might be relevant during the start-up period of large nuclear power plants, as well as during transition phases between steady operation states.

The authors propose to perform a detailed study of the long-time count rate and temperature behavior at a different TRIGA Mk II reactor, since it is not possible with the 7 hours per day, 5 days per week operating schedule at TRIGA Center Atominstitut. 
Furthermore, precise temperature measurements of the cooling water across multiple positions inside the reactor pool, the fuel elements, the graphite reflector, and the thermal column are suggested. 
Different reactor operation modes with the goal of independently influencing the temperature of the fuel elements, cooling water and graphite reflector could allow further insights into this complex coupled system (e.g. by operating the reactor at a low power level of a few \unit{\kilo\watt} and manually heating the water). 
Additionally, it should be investigated, if the apparent count rate reduction could result from a shift of the thermal neutron energy spectrum towards slightly higher energies, reducing the detection efficiency of the ${}^3$He detectors. 
Thus, a time-dependent investigation of the thermal neutron energy spectrum at different beam tubes via time of flight experiments would be highly valuable.
Validating the current observations with one of the many already available simulation tools, coupling neutron physics with CFD calculations (as mentioned in \cref{sec:introduction}), would be tremendously beneficial. 
A comprehensive investigation of the effect would be important for the broader reactor physics, technology and engineering community, extending beyond small research reactors, warranting additional research investments. 
 
\section{Declaration of Interests}
\paragraph{Research support}
The authors acknowledge TU Wien Bibliothek for financial support through its Open Access Funding Programme. This research was funded in part by the Austrian Science Fund (FWF) (Grant DOI 10.55776/PAT3585625 and 10.55776/I5427). 
\paragraph{Relationships}
There are no additional relationships to disclose.
\paragraph{Patents and Intellectual Property}
There are no patents to disclose.
\paragraph{Other Activities}
There are no additional activities to disclose.

\section{Author Contributions}
\textbf{Sebastian Dorer:} Conceptualization, Methodology, Validation, Formal analysis, Investigation, Resources, Data Curation, Writing - Original Draft, Writing - Review \& Editing, Visualization, Supervision, Project administration. 
\textbf{Clemens Trunner:} Conceptualization, Methodology, Validation, Formal analysis, Investigation, Resources, Data Curation, Writing - Original Draft, Writing - Review \& Editing, Visualization, Supervision, Project administration. 
\textbf{Erwin Jericha:} Conceptualization, Methodology, Validation, Resources, Writing - Review \& Editing, Supervision, Project administration. 
\textbf{Stephan Sponar:} Conceptualization, Methodology, Validation, Resources, Writing - Review \& Editing, Supervision, Project administration. 
\textbf{Robert Bergmann:} Resources.
\textbf{Dieter Hainz:} Conceptualization, Resources.
\textbf{Thomas Stummer:} Conceptualization, Resources.
\textbf{Mario Villa:} Conceptualization, Resources, Supervision.

\printbibliography
\end{document}